# Shell Model Calculations for Proton-rich Zn Isotopes via New Generated Effective Interaction by Artificial Neural Networks

Serkan Akkoyun

*Department of Physics, Faculty of Science, Sivas Cumhuriyet University, Sivas, Turkey*

**Abstract:** In this study, the artificial neural network method has been employed for the generation of the new two-body matrix elements which is used for pfg shell nuclei. For this purpose, jj44b interaction Hamiltonian has been considered as a source. After the generation of the new Hamiltonian, both, original and new generated, are tested on proton-rich Zn isotopes. According to the results, the calculated values are close to the each other. As well the results from new interaction (jj44b_nn) are closer to the available experimental values in some cases.

## 1.INTRODUCTION

Forthe investigation of nuclear structure, nuclear shell model is a quite successful. Similar to the more familiar model of electronic shell model of atom, the nucleons are considered in the shells in the nucleus. If the number of nucleons equals to 2, 8, 20 28, 50, 82 or 126 which is named as magic number, the shells are said to be closed. The nuclei whose only one type nucleon number is magic are called magic nuclei. In the case of both types are magic, it is called as doubly-magic nuclei. In nuclear shell model (SM) calculations, a double-magic nucleus is considered as an inert core. According to the double-magic nuclei, there are different shell groups. The core nucleons do not interact with each other or the nucleons outside the core. Only valance nucleons above the closed shell group are taken in the calculations. The space of the valance nucleons is model space.

The interactions of the nucleons in the model space with each other or the core are obtained via SM calculations [1-7]. Therefore, the correct estimations of the physical results in SM model, determination of the interaction Hamiltonian plays crucial role. In this paper, we have generated new interaction matrix elements from existed jj44b interaction [8]. This interaction is used for SM calculations of the nuclei outside the $^{56}$Ni double-magic core and includes 133 two-body interaction matrix elements. The model space is pfg shell and contains the nuclei from mass number A=56 to 100. For the generation, we have applied artificial neural network (ANN) method which mimics the human brain function. Recently, ANN has been used in many fields of nuclear physics [9,13]. After generation of the new interaction matrix elements, the $2^+$, $4^+$ energies, R4/2 ratio, B(E2) values for proton-rich Zn isotopes have been calculated in the scope of SM. Kshell shell model code has been used which enables to perform nuclear shell-model calculations with M-scheme representation with the thick-restart Lanczos method [14].

The paper is organized fallows. In section 1, a brief introduction is given to the subject. In section 2, ANN method and shell model calculations have been summarized. Results of the ANN and the SM calculations for existed jj4b interaction and newly generated jj44b_nn interaction for Zn isotopes have been given in comparison with the available literature values in section 3. Finally, in section 4, discussions on the obtained results have been given with conclusions.

## 2. MATERIAL and METHODS

*2.1 Artificial Neural Network (ANN) Method*

A mathematical tool ANN mimics the human brain functionality [15]. The structure of the ANN is composed of several processing units which are called neurons. The neurons are located in different layers in groups and connected each other via adaptive synaptic weights. The first group of neurons is in input layer and receives data from outside. The data is transmitted to next group of neurons in hidden layer by weights. Finally, the processed data flows to the output layer neurons. In our calculation, we have used feed-forward ANN with four layers in order to generate new two-body interaction matrix elements (tbme). The input layer consists of six neurons corresponding to a, b, c, d, J and T. The a, b, c, d values are the single particle orbits of the model space nucleons, J and T are total paired angular momentum and isospin, respectively. Due to the giving best results among the others, two hidden layers with 6 and 9 neurons in each are used as optimized value. The output layer neuron corresponds to the tbme. Therefore, the architecture of ANN is 6-6-9-1 and the total numbers of adjustable weights between neurons are 99 in this work. The hidden neuron activation function is tangent hyperbolic which is a sigmoid-like function generally used in the ANN calculations. For details of the ANN, the reader is referred to Haykin [15].

*2.2 Shell Model Calculation*

In the SM, valance nucleons move in a finite number of j-orbits and their Hamiltonian of the valance nucleons is given by

$$H = E_0 + \sum_i \varepsilon_i + \sum_{a,b,c,d} <ab;JT|V|cd;JT>$$

where $E_0$ is the energy of the inert core, $\varepsilon_i$ is single particle energies (spe) of the valance orbits and the last term *<ab;JT|V|cd;JT>* is two-body residual interaction among the valance particles. In the calculations with jj44bor jj44b_nn effective interactions, spe values are -9.2859, -9.6566, -8.2695 and -5.8944 for $p_{3/2}$, $f_{5/2}$, $p_{1/2}$ and $g_{9/2}$ shells. The valance nucleons are distributed in this pfg shells (Fig.1). The interaction Hamiltonian is defined by a set of 133 tbme. We have considered double magic $^{56}$Ni isotope as a core whose proton and neutron numbers are 28. The nucleons in the core with J=0 do not move from the core.

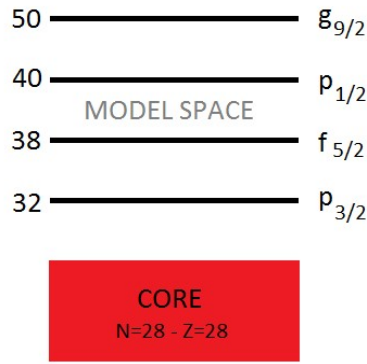

**Fig. 1** The pfg model space above the $^{56}$Ni core for the calculations of Zn isotopes

There are many available codes in the literature which is written for the shell model calculations such as NuShell [16], Redstick [17], Bigstick [18], Antoine [19], Oxbash [20]. For the SM calculations for nuclear properties of proton-rich Zn isotopes, Kshell [14] computer code has been used. The code is a powerful computer code in order to calculate the energy levels, spins/parities, electric/magnetic quadrupole moments, electric/magnetic transition probabilities and one-particle spectroscopic factors in the nuclei. The code is used on a Linux operating system with a many-core CPU and OpenMP library. It is also used on a parallel computer with hybrid MPI+OpenMP parallel programming. If enough memory is available on the computers, up to tens of billions M-scheme dimension is capable.

## 3. RESULTS AND DISCUSSIONS

In the first stage of the study, we have employed ANN to generate new interaction matrix elements from an existing data for the shell model calculations of pfg shell nuclei. By using constructed ANN mentioned in Section 2.1, we have obtained the elements whose values are not so far from the original ones. As can be seen in Table 1 that, the maximum difference between original and generated elements is 0.4861 for off-diagonal *<41;50|V|42;50>* tbme. The minimum value is 0 for several tbme. The more the hidden layer neurons the closest the result between original and neural network data is. For the hidden neuron numbers 8-10, the constructed ANN starts to memorize the data and the generated tbme becomes to be same as the original tbme. Although the values are close each other for used ANN with 6-9 hidden neurons, the results of SM calculations have been improved by using new generated tbme. In order to see this improvement, SM calculations have been performed on proton-rich even-even Zn isotopes.

In Table 2, we have given the $2^+$ and $4^+$ energies of $^{58-62}$Zn isotopes. For $^{58}$Zn, the theoretical calculated $2^+$ energy values are same from jj44b and jj44b_nn. For $^{60}$Zn isotope, jj44b_nn gives results closer to the experimental value whereas for $^{62}$Zn isotope, jj44b gives closer results. The calculations of $4^+$ energy levels, jj44b_nn is better both for $^{58}$Zn and $^{60}$Zn.

In Table 3, we have shown the B(E2) values for ground state to first $2^+$ excited state and from $2^+$ excited state to ground state. Only available experimental or adopted literature value is for ground state to $2^+$ state in $^{62}$Zn isotope. As is clear in the table that the jj44b_nn is slightly closer to the experimental value. For the other Zn isotopes, the results are very close to each other. In Table 4, we have listed the B(E2) values for first $2^+$ excited state to first $4^+$ excited state and from $4^+$ excited state to $2^+$ excited state. There is no available experimental or adopted value in the literature for these values.

**Table 1.** Two-body matrix elements of jj44b and newly generated jj44b_nn interactions. The orbits are labeled by 1=$p_{3/2}$, 2=$f_{5/2}$, 3=$p_{1/2}$ and 4=$g_{9/2}$ single particle orbitals.

| a | b | c | d | J | T | jj44b | jj44b_nn | a | b | c | d | J | T | jj44b | jj44b_nn |
|---|---|---|---|---|---|-------|----------|---|---|---|---|---|---|-------|----------|
| 1 | 1 | 1 | 1 | 1 | 0 | -0,8113 | -0,8081 | 4 | 4 | 4 | 4 | 9 | 0 | -2,1260 | -2,1283 |
| 1 | 1 | 1 | 1 | 3 | 0 | -0,4985 | -0,5162 | 1 | 1 | 1 | 1 | 0 | 1 | -1,3259 | -1,2921 |
| 1 | 1 | 1 | 1 | 5 | 0 | -2,1319 | -2,1179 | 1 | 1 | 1 | 1 | 2 | 1 | -0,3349 | -0,3358 |
| 1 | 1 | 2 | 1 | 1 | 0 | 0,5205 | 0,5580 | 1 | 1 | 1 | 1 | 4 | 1 | 0,3138 | 0,3728 |
| 1 | 1 | 2 | 1 | 3 | 0 | 0,3487 | 0,3125 | 1 | 1 | 2 | 1 | 2 | 1 | -0,2903 | -0,2349 |
| 1 | 1 | 3 | 1 | 3 | 0 | -0,6053 | -0,6000 | 1 | 1 | 2 | 1 | 4 | 1 | -0,4315 | -0,4270 |
| 1 | 1 | 2 | 2 | 1 | 0 | 0,1736 | 0,1988 | 1 | 1 | 3 | 1 | 2 | 1 | -0,7133 | -0,6980 |
| 1 | 1 | 2 | 2 | 3 | 0 | 0,0320 | 0,0320 | 1 | 1 | 2 | 2 | 0 | 1 | -0,7430 | -0,8444 |
| 1 | 1 | 3 | 2 | 1 | 0 | 0,1136 | -0,0226 | 1 | 1 | 2 | 2 | 2 | 1 | 0,0098 | 0,0499 |
| 1 | 1 | 3 | 3 | 1 | 0 | -0,2385 | -0,2908 | 1 | 1 | 3 | 2 | 2 | 1 | 0,5004 | 0,4537 |
| 1 | 1 | 4 | 4 | 1 | 0 | -0,8790 | -0,8191 | 1 | 1 | 3 | 3 | 0 | 1 | -0,4773 | -0,4915 |
| 1 | 1 | 4 | 4 | 3 | 0 | -0,3107 | -0,2993 | 1 | 1 | 4 | 4 | 0 | 1 | 1,8722 | 1,8353 |
| 1 | 1 | 4 | 4 | 5 | 0 | -0,1703 | -0,1798 | 1 | 1 | 4 | 4 | 2 | 1 | 0,2798 | 0,4341 |
| 2 | 1 | 2 | 1 | 1 | 0 | -1,9735 | -1,9779 | 1 | 1 | 4 | 4 | 4 | 1 | 0,1879 | 0,0334 |
| 2 | 1 | 2 | 1 | 2 | 0 | -1,2445 | -1,2437 | 2 | 1 | 2 | 1 | 1 | 1 | 0,2446 | 0,2890 |
| 2 | 1 | 2 | 1 | 3 | 0 | -0,6597 | -0,6621 | 2 | 1 | 2 | 1 | 2 | 1 | 0,4202 | 0,3899 |
| 2 | 1 | 2 | 1 | 4 | 0 | -1,6087 | -1,6137 | 2 | 1 | 2 | 1 | 3 | 1 | 0,5865 | 0,6269 |
| 2 | 1 | 3 | 1 | 2 | 0 | -0,7312 | -0,6938 | 2 | 1 | 2 | 1 | 4 | 1 | -0,1017 | -0,1027 |
| 2 | 1 | 3 | 1 | 3 | 0 | 0,6758 | 0,6572 | 2 | 1 | 3 | 1 | 2 | 1 | -0,4233 | -0,3843 |
| 2 | 1 | 2 | 2 | 1 | 0 | 0,0665 | 0,0659 | 2 | 1 | 3 | 1 | 3 | 1 | 0,0290 | 0,0455 |
| 2 | 1 | 2 | 2 | 3 | 0 | 0,3863 | 0,4129 | 2 | 1 | 2 | 2 | 2 | 1 | -0,0657 | -0,1104 |
| 2 | 1 | 3 | 2 | 1 | 0 | 0,6218 | 0,6408 | 2 | 1 | 3 | 2 | 1 | 1 | 0,0725 | 0,0584 |
| 2 | 1 | 3 | 2 | 2 | 0 | -0,3687 | -0,4616 | 2 | 1 | 3 | 2 | 2 | 1 | -0,0633 | -0,1426 |
| 2 | 1 | 3 | 3 | 1 | 0 | 0,5089 | 0,6585 | 2 | 1 | 4 | 4 | 2 | 1 | 0,5475 | 0,5788 |
| 2 | 1 | 4 | 4 | 1 | 0 | 0,3037 | 0,2474 | 2 | 1 | 4 | 4 | 4 | 1 | 0,5235 | 0,5430 |
| 2 | 1 | 4 | 4 | 3 | 0 | -0,0341 | -0,0234 | 3 | 1 | 3 | 1 | 2 | 1 | -0,1648 | -0,2051 |
| 3 | 1 | 3 | 1 | 2 | 0 | -0,4481 | -0,4367 | 3 | 1 | 3 | 1 | 3 | 1 | 0,6638 | 0,6220 |
| 3 | 1 | 3 | 1 | 3 | 0 | -1,4983 | -1,4898 | 3 | 1 | 2 | 2 | 2 | 1 | -0,2554 | -0,2632 |
| 3 | 1 | 2 | 2 | 3 | 0 | -0,0006 | 0,0032 | 3 | 1 | 3 | 2 | 2 | 1 | 0,4423 | 0,4458 |
| 3 | 1 | 3 | 2 | 2 | 0 | -0,5797 | -0,6143 | 3 | 1 | 4 | 4 | 2 | 1 | 0,6139 | 0,5485 |
| 3 | 1 | 4 | 4 | 3 | 0 | -0,1910 | -0,1704 | 4 | 1 | 4 | 1 | 2 | 1 | -0,6032 | -0,6032 |
| 4 | 1 | 4 | 1 | 2 | 0 | -3,5289 | -3,5389 | 4 | 1 | 4 | 1 | 3 | 1 | -0,0920 | -0,0419 |
| 4 | 1 | 4 | 1 | 3 | 0 | -1,8098 | -1,8147 | 4 | 1 | 4 | 1 | 4 | 1 | 0,4452 | 0,2406 |
| 4 | 1 | 4 | 1 | 4 | 0 | -1,2352 | -1,2256 | 4 | 1 | 4 | 1 | 5 | 1 | 0,2804 | 0,4718 |
| 4 | 1 | 4 | 1 | 5 | 0 | -1,4659 | -1,4834 | 4 | 1 | 4 | 1 | 6 | 1 | 0,6757 | 0,6048 |
| 4 | 1 | 4 | 1 | 6 | 0 | -0,8493 | -0,8040 | 4 | 1 | 4 | 1 | 7 | 1 | -0,7200 | -0,7734 |
| 4 | 1 | 4 | 1 | 7 | 0 | -2,3659 | -2,4047 | 4 | 1 | 4 | 2 | 3 | 1 | -0,5352 | -0,4555 |
| 4 | 1 | 4 | 2 | 3 | 0 | -0,8169 | -0,7536 | 4 | 1 | 4 | 2 | 4 | 1 | -0,1519 | -0,2064 |
| 4 | 1 | 4 | 2 | 4 | 0 | 0,5634 | 0,2266 | 4 | 1 | 4 | 2 | 5 | 1 | -0,0907 | 0,0130 |
| 4 | 1 | 4 | 2 | 5 | 0 | -0,1897 | 0,2964 | 4 | 1 | 4 | 2 | 6 | 1 | -0,4143 | -0,3692 |
| 4 | 1 | 4 | 2 | 6 | 0 | 0,8527 | 0,6543 | 4 | 1 | 4 | 3 | 4 | 1 | 0,1336 | 0,1116 |
| 4 | 1 | 4 | 3 | 4 | 0 | -0,8072 | -0,6775 | 4 | 1 | 4 | 3 | 5 | 1 | -0,0729 | -0,0952 |
| 4 | 1 | 4 | 3 | 5 | 0 | -0,8831 | -0,9912 | 2 | 2 | 2 | 2 | 0 | 1 | -0,7572 | -0,7504 |
| 2 | 2 | 2 | 2 | 1 | 0 | -0,8551 | -0,8573 | 2 | 2 | 2 | 2 | 2 | 1 | 0,0862 | 0,1305 |
| 2 | 2 | 2 | 2 | 3 | 0 | -2,0525 | -2,0635 | 2 | 2 | 3 | 2 | 2 | 1 | 0,2157 | 0,2559 |
| 2 | 2 | 3 | 2 | 1 | 0 | -1,3769 | -1,3686 | 2 | 2 | 3 | 3 | 0 | 1 | -1,1326 | -1,1276 |
| 2 | 2 | 3 | 3 | 1 | 0 | 0,6256 | 0,6214 | 2 | 2 | 4 | 4 | 0 | 1 | 0,9686 | 0,9654 |
| 2 | 2 | 4 | 4 | 1 | 0 | 0,5998 | 0,5808 | 2 | 2 | 4 | 4 | 2 | 1 | 0,5567 | 0,4995 |
| 2 | 2 | 4 | 4 | 3 | 0 | 0,4825 | 0,5337 | 3 | 2 | 3 | 2 | 1 | 1 | 0,3052 | 0,3135 |
| 3 | 2 | 3 | 2 | 1 | 0 | -2,3068 | -2,3085 | 3 | 2 | 3 | 2 | 2 | 1 | -0,3941 | -0,4021 |
| 3 | 2 | 3 | 2 | 2 | 0 | -2,0031 | -1,9665 | 3 | 2 | 4 | 4 | 2 | 1 | -0,1474 | -0,0989 |
| 3 | 2 | 3 | 3 | 1 | 0 | -0,3608 | -0,3664 | 4 | 2 | 4 | 2 | 3 | 1 | -0,8576 | -0,8774 |
| 3 | 2 | 4 | 4 | 1 | 0 | 0,5368 | 0,5290 | 4 | 2 | 4 | 2 | 4 | 1 | 0,1169 | 0,1576 |
| 4 | 2 | 4 | 2 | 3 | 0 | -0,9996 | -1,0071 | 4 | 2 | 4 | 2 | 5 | 1 | -0,1975 | -0,2201 |
| 4 | 2 | 4 | 2 | 4 | 0 | -0,8777 | -0,8467 | 4 | 2 | 4 | 2 | 6 | 1 | 0,8224 | 0,7916 |
| 4 | 2 | 4 | 2 | 5 | 0 | -0,3781 | -0,4948 | 4 | 2 | 4 | 3 | 4 | 1 | 0,2570 | 0,2396 |
| 4 | 2 | 4 | 2 | 6 | 0 | -2,3615 | -2,3469 | 4 | 2 | 4 | 3 | 5 | 1 | -0,4919 | -0,4556 |
| 4 | 2 | 4 | 3 | 4 | 0 | 1,0058 | 0,9069 | 3 | 3 | 3 | 3 | 0 | 1 | -0,1139 | -0,1060 |
| 4 | 2 | 4 | 3 | 5 | 0 | -0,3069 | -0,1848 | 3 | 3 | 4 | 4 | 0 | 1 | 0,8110 | 0,8110 |
| 3 | 3 | 3 | 3 | 1 | 0 | -1,0199 | -1,0155 | 4 | 3 | 4 | 3 | 4 | 1 | 0,3293 | 0,3363 |
| 3 | 3 | 4 | 4 | 1 | 0 | -0,2545 | -0,2517 | 4 | 3 | 4 | 3 | 5 | 1 | -0,2067 | -0,2205 |
| 4 | 3 | 4 | 3 | 4 | 0 | -1,6832 | -1,7087 | 4 | 4 | 4 | 4 | 0 | 1 | -1,4086 | -1,4040 |
| 4 | 3 | 4 | 3 | 5 | 0 | -1,1173 | -1,1116 | 4 | 4 | 4 | 4 | 2 | 1 | -1,0366 | -1,0421 |
| 4 | 4 | 4 | 4 | 1 | 0 | -1,0265 | -1,0103 | 4 | 4 | 4 | 4 | 4 | 1 | -0,2288 | -0,2259 |
| 4 | 4 | 4 | 4 | 3 | 0 | -0,5628 | -0,5201 | 4 | 4 | 4 | 4 | 6 | 1 | 0,2066 | 0,1920 |
| 4 | 4 | 4 | 4 | 5 | 0 | -0,5590 | -0,6344 | 4 | 4 | 4 | 4 | 8 | 1 | 0,2457 | 0,2489 |
| 4 | 4 | 4 | 4 | 7 | 0 | -0,8680 | -0,8583 | | | | | | | | |

**Table 2.** First 2+ and 4+ energies of proton rich Zn isotopes

|  | 2⁺ Energy (keV) | | | 4⁺ Energy (keV) | | |
|---|---|---|---|---|---|---|
| *Isotope* | *exp* | *jj4b* | *jj4b_nn* | *exp* | *jj4b* | *jj4b_nn* |
| ⁵⁸Zn | 1356 | 1683 | 1683 | 2499 | 2299 | 2323 |
| ⁶⁰Zn | 1004 | 998 | 1001 | 2193 | 2112 | 2139 |
| ⁶²Zn | 954 | 958 | 960 | 2186 | 2220 | 2238 |

**Table 3.** B(E2) values between ground state and first 2+ state for proton rich Zn isotopes

|  | B(E2;0-2) | | | B(E2;2-0) | | |
|---|---|---|---|---|---|---|
| *Isotope* | *exp* | *jj4b* | *jj4b_nn* | *exp* | *jj4b* | *jj4b_nn* |
| ⁵⁸Zn | - | 466.87 | 467.41 | - | 93.37 | 93.48 |
| ⁶⁰Zn | - | 709.39 | 710.86 | - | 141.88 | 142.17 |
| ⁶²Zn | 1240 | 807.99 | 809.55 | - | 161.60 | 161.91 |

**Table 4.** B(E2) values between first 2+ state and 4+ state for proton rich Zn isotopes

|  | B(E2;2-4) | | | B(E2;4-2) | | |
|---|---|---|---|---|---|---|
| *Isotope* | *exp* | *jj4b* | *jj4b_nn* | *exp* | *jj4b* | *jj4b_nn* |
| ⁵⁸Zn | - | 155.68 | 155.52 | - | 86.49 | 86.40 |
| ⁶⁰Zn | - | 308.5 | 309.08 | - | 171.39 | 171.71 |
| ⁶²Zn | - | 373.89 | 375.27 | - | 207.72 | 208.49 |

We have also calculated $R_{4/2}$ and B(E2; 4->2)/B(E2;2->0) ratios for Zn isotopes. In Fig.2, we have shown the $R_{4/2}$ ratios from experimental results and theoretical results by jj44b and jj44b_nn interactions. It is clear in figure that, the results from newly generated jj44b_nn tbme are slightly closer to the experimental data for ⁵⁸⁻⁶⁰Zn isotopes.

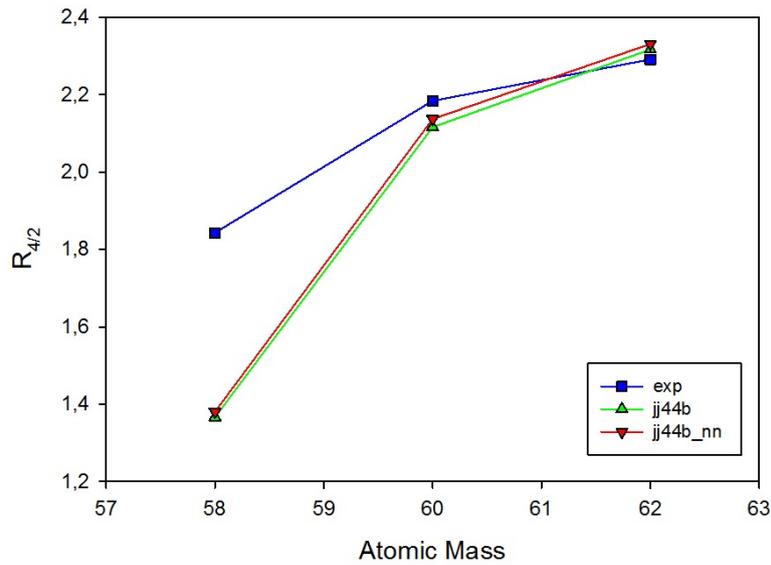

**Fig. 2** $R_{4/2}$ values for proton-rich Zn isotopes.

Finally in Fig.3, we have shown the B(E2; 4->2)/B(E2;2->0) ratios from experimental results and theoretical results by jj44b and jj44b_nn interactions. Only available experimental values are for $^{62}$Zn isotope with the value of 1.55. Neither jj44b nor jj44b_nn results are close to this value. The results from both calculations is almost same each other.

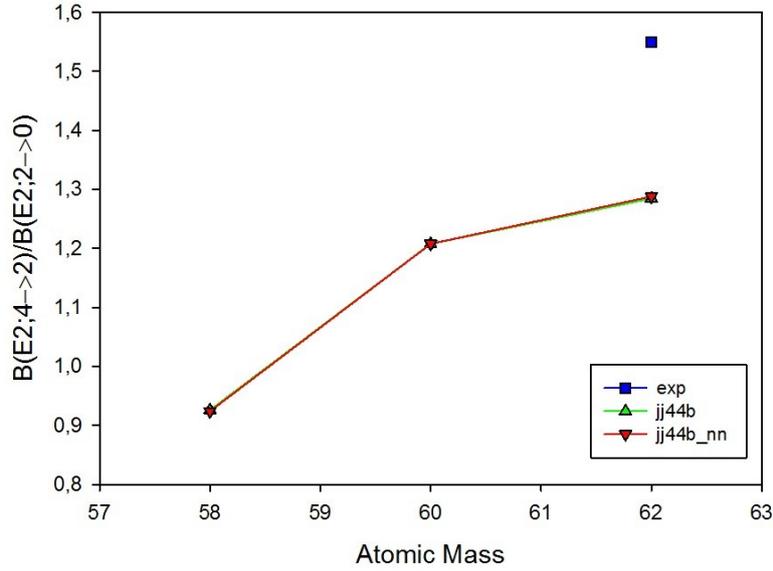

**Fig. 3** B(E2; 4->2)/B(E2;2->0) values for proton-rich Zn isotopes.

## 5. ACKNOWLEDGEMENTS

This work is supported by the scientific research project fund of Sivas Cumhuriyet University under the project number SHMYO-015.